\documentclass[12pt]{iopart}
\usepackage[latin9]{inputenc}
\usepackage{hyperref}
\usepackage{array}
\usepackage{float}
\usepackage{booktabs}
\usepackage{textcomp}
\usepackage{multirow}
\usepackage{amstext}
\usepackage{graphicx}
\usepackage{cite}
\makeatletter

\expandafter\let\csname equation*\endcsname=\relax

\expandafter\let\csname endequation*\endcsname=\relax

\usepackage{amsmath,amssymb,amsthm}

\makeatother

\begin{document}
\title[]{
Emergent $\rm O(4)$ symmetry at an one-dimensional
deconfined quantum tricritical point}
\author{Ning Xi}
\address{Department of Physics and Beijing Key Laboratory of Opto-electronic
Functional Materials and Micro-nano Devices, Renmin University of
China, Beijing 100872, China}
\author{Rong Yu}
\address{Department of Physics and Beijing Key Laboratory of Opto-electronic
Functional Materials and Micro-nano Devices, Renmin University of
China, Beijing 100872, China}
\ead{rong.yu@ruc.edu.cn}
\begin{abstract}
We show an $\rm O(4)$ symmetry emerges at a deconfined quantum tricritical point of a valence bond solid and two ferromagnetic phases in an $S = 1/2$ frustrated spin chain by combining analytical analysis and numerical calculations with the time evolution of infinite matrix product states. With this symmetry the valence-bond solid and the three magnetic order parameters form an $\rm O(4)$ pseudovector in the infrared limit, and can continuously rotate into each other. We numerically determine the location of the quantum tricritical point and study the scaling of the correlation functions of the $\rm O(4)$ vector components and associated conserved currents. The critical behaviors of these correlation functions are all in accord with field theoretical results. The emergent $\rm O(4)$ symmetry at the tricritical point is justified by the integer value of the scaling dimension of the emergent Noether conserved currents. Our findings not only give direct evidence of such a high emergent symmetry at an one-dimensional valence bond solid to magnetic transition but also shed light on exploring emergent symmetries in higher dimensions.
\end{abstract}
\vspace{2pc}
\maketitle

\section{Introduction}

Quantum phase transitions (QPTs) in spin systems are usually accompanied by abundant
emergent phenomena, including unconventional spin dynamics, fractionalized excitations,
topological excitations, and enhanced symmetry~\cite{sachdev2011BOOK,giamarchi2004quantum,fradkin2013BOOK,chaikin1995BOOK,tsvelik2007BOOK,nagaosa1999BOOK,lacroix2011BOOK,diep2013BOOK,balents2010Nature,Shao2016,Coldea2010,SCZhang1997science,Nahum2015,DQCP2004}.
Among these, the origin of emergent enhanced continuous symmetry at
the transition point is one intriguing theme that has been extensively
investigated\cite{Nahum2015,Coldea2010,SCZhang1997science,DQCP2004,Gazit2018PNAS,zhao2019symmetry,HaiyuanE82021,CuiYSCVO2019,Lee2019DMRGSSL,Cui2021,xi2021firstorder}.

As one promising theory, the theory of deconfined quantum critical
point (DQCP) successfully predicts the existence of emergent
continuous symmetry at the continuous transition between two
spontaneous symmetry breaking (SSB) phases in two-dimensional (2D) systems\cite{DQCP2004,Senthil2004,Sadvik2007,Deng2013}.
At the DQCP, besides the emergent deconfined fractionalized spin excitations, the enhanced
symmetry allows continuous rotation between the order parameters of
the two ordered phases\cite{Meng2019,PhysRevX031052,Meng2018,Meng2017}.
For example, an emergent $\rm SO(5)$ symmetry has been observed numerically at the DQCP between the antiferromagnetic (AFM) and valence bond solid (VBS) states on the 2D square lattice\cite{Nahum2015,Gazit2018PNAS,WangChong2017PhysRevX}.

The development of the DQCP theory sheds light on understanding the emergent symmetry at the VBS to magnetic transition. However, it is sufficient but not necessary for the emergent symmetry at a phase transition.
For instance, evidences of enhanced continuous symmetry at first-order transitions between AFM and VBS phases are recently reported~\cite{zhao2019symmetry,Cui2021}. In these systems the origin of the emergent symmetry remain elusive.

Despite the complex situation in high dimensional systems, the emergent continuous symmetry in 1D systems is more accessible. This is not only because more controllable analytical and numerical tools are available, but also because the significant quantum fluctuations in 1D systems constrains the type of transition.
This is especially true for the systems where the Lieb-Schultz-Mattis (LSM) theorem holds. For example, in an $S$ = 1/2 chain with spin rotational and lattice translational symmetries, the ground state either breaks
the translational symmetry to form a VBS state or keeps plainly gapless with the symmetries preserved~\cite{LIEB1961407}.
Therefore, for a LSM system with spin rotational and lattice translational
symmetries, a first-order transition with an enhanced continuous symmetry can not take place, and the ground state at the transition point, being gapless, least preserves the symmetries of
the underlying Hamiltonian, and may even extend to an enhanced symmetry~\cite{LIEB1961407,LSM2015,LSM2017}. In other words, the DQCP theory applies to these systems in a more universal way. Recently, some works studied possible DQCP in 1D spin systems and
obtained some interesting results.\cite{Huang2019,Huang2020,Toshiya2019,MotrunichNUM2019,MotrunichTHE2019,Furusaki2012,Sandvik2018Heisenberg,mudry2019quantum,PhysRevE.104.064121}.

\begin{figure}
\includegraphics[width=0.6\textwidth]{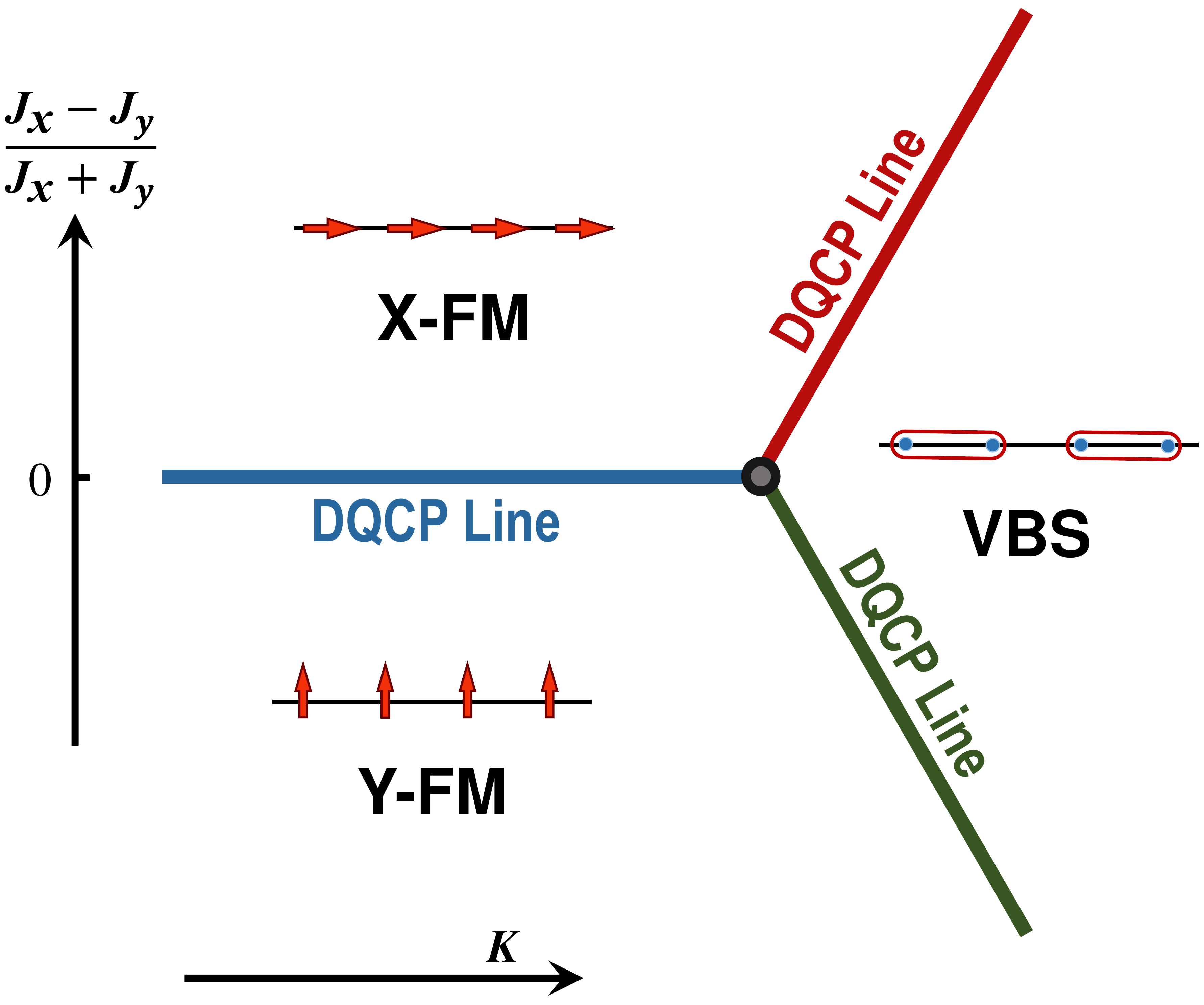}
\centering
\caption{A schematic phase diagram for the spin model defined in Eq. \ref{eq:spinmodel}.
The isotropic DQCP line with microscopic ${\rm U(1)}$ symmetry (blue line)
separates the $X$-direction and $Y$-direction FM phases. The DQCP
lines between the FM and VBS phases (red and green line) with emergent
${\rm O(2)}\text{\ensuremath{\times}}{\rm O(2)}$ symmetry have been studied in
Ref. \cite{Huang2019}. As shown later in this paper, an enhanced ${\rm O(4)}$ symmetry emerges at the intersection point of these three quantum critical lines, which is a quantum tricritical point (black point). A more completed phase diagram is described in Ref. \cite{MotrunichNUM2019}.
\label{fig:1}}
\end{figure}

To clarify the nature of the emergent continuous symmetry accompanied by the magnetic-VBS transition in 1D systems, we investigate the QPT of a frustrated spin-1/2 chain. Inspired by the results of an spin-1/2 AFM Heisenberg chain, where the microscopic $\rm SU(2)$ symmetry extends to an $\rm SO(4)$ symmetry in the continuous limit\cite{Sandvik2018Heisenberg,yang2020deconfined,Affleck1985PRL,Affleck1985PRB,tsvelik2007BOOK}, we show that an $\rm O(4)$ symmetry emerges at the quantum tricritical point of the VBS and two ferromagnetic (FM) phases in our model. To be specific, the Hamiltonian of our model reads
\begin{equation}
H = \sum_{i} \left(-J_{x}S_{i}^{x}S_{i+1}^{x}-J_{y}S_{i}^{y}S_{i+1}^{y}\right) +K\left(S_{i}^{x}S_{i+2}^{x}+S_{i}^{y}S_{i+2}^{y}\right).\label{eq:spinmodel}
\end{equation}
Here $J_{x}$ and $J_{y}$ refer to nearest-neighbor (n.n.) FM exchange interactions
and $K$ is the next n.n. AFM exchange coupling. The model supports
an $X$-direction FM, a $Y$-direction FM, and a VBS in the ground-state
phase diagram, as depicted in Fig. \ref{fig:1}.

For $J_{x}>J_{y}$ and small $K$, the ground state is a FM with spins
ordered along the $X$-direction($X$-FM). Likewise, the ground state
is a $Y$-FM for $J_{x}<J_{y}$ and small $K$. The $X$-FM and $Y$-FM
states meet at the isotropic line $J_{x}=J_{y}$ (blue line in Fig.
\ref{fig:1}) where the spin rotational symmetry of the Hamiltonian
is enhanced from ${\rm Z_{2}}^{x}\times {\rm Z_{2}}^{y}$ to ${\rm U(1)}$. Across this
isotropic line by tuning the parameter $J_{x}-J_{y}$ in the model,
the system undergoes a continuous transition, i.e. a line of DQCPs\cite{Ning2022}.
For sufficiently large $K$, the ground state is a VBS that breaks
the translational symmetry. It has been proposed that DQCPs with emergent
${\rm O(2)}\times {\rm O(2)}$ symmetry\cite{Huang2019} exist along the FM and
VBS phase boundary (red and green lines in Fig. \ref{fig:1}).

The two ${\rm O(2)}$ symmetries between $X$-FM($Y$-FM) and VBS respectively
correspond to the rotation between the order parameters of VBS
and $X$-FM ($Y$-FM) and the rotation between the order parameters
of $Y$-FM ($X$-FM) and $Z$-AFM. Meanwhile, the ${\rm U(1)}$ symmetry
between $X$-FM and $Y$-FM guarantees the continuous rotation between
these two order parameters. As a consequence, at the tricritical point (black point in Fig. \ref{fig:1}),
the four order parameters of $X$-FM, $Y$-FM, $Z$-AFM, and VBS are expected
to be rotated into each other by an enhanced ${\rm O(4)}$ symmetry.

In this work, we present results of analytical analysis and numerical calculations to show an $\rm O(4)$
symmetry indeed emerges in the infrared limit at the tricritical point of the $X$-FM, $Y$-FM, and VBS phases.
The rest of this paper is organized as follows. In Sec. \ref{sec:Effective-Models},
we review the field theoretical description of this 1D model, from which the emergent $\rm O(4)$ symmetry is expected.
In Sec. \ref{sec:Numerical-Methods}, we introduce our numerical technique based on the time evolution of infinite matrix product states (MPS) for calculating the correlation functions of ${\rm O(4)}$ vector components and
associated conserved currents from which the critical exponent $\eta$ (related to scaling dimensions $\Delta$) for each component is extracted. Then in Sec.
\ref{sec:Numerical-Analysis-of}, we numerically study the transition
along the isotropic line and determine the exact location of the tricritical
point. In Sec. \ref{sec:Numerical-evidence-of}, the emergent ${\rm O(4)}$
symmetry and the associated conserved currents correlation at the
tricritical point are investigated in detail. We show that the scaling dimensions of all the emergent Noether conserved currents associated with the $\rm O(4)$ vector are pinned to an integer of $\Delta=1$, which justifies the existence of the emergent ${\rm O(4)}$ symmetry.

\section{Low-energy effective model description\label{sec:Effective-Models}}

The model of Eq.~\eqref{eq:spinmodel} has been analyzed by using several different approaches\cite{Huang2019,Huang2020,MotrunichNUM2019,MotrunichTHE2019}.
Here we adopt a direct bosonization approach which can give a clear understanding
of the QPT and emergent symmetries. Following the standard procedure of bosonization~\cite{MotrunichTHE2019,tsvelik2007BOOK}, the low-energy symmetry-preserving bosonized hamiltonian reads as
\begin{equation}
H_{b}=H_{\mathrm{0}}+H_{\mathrm{int}}, \label{eq:bosonization}
\end{equation}
where
\begin{equation}
H_{0}=\frac{v}{2} \int\mathrm{d}r \left[\frac{1}{g}\left(\partial_{r}\Theta\right)^{2} +g\left(\partial_{r}\Phi\right)^{2}\right], \label{eq:bosonizationH0}
\end{equation}
and
\begin{equation}
H_{\mathrm{int}}=\int d\tau dr\left[\lambda_{\Theta}\cos(4\Theta)+\lambda_{\Phi}\cos(2\Phi)\right].\label{eq:bosonizationHint}
\end{equation}
Here $v$ is the Fermi velocity, and $g$ is the Luttinger parameter. $\lambda_{\Theta}$ and $\lambda_{\Phi}$
are two interaction couplings that can drive the ground state to different orders. $\Theta$ and $\Phi$ are two dual bosonic fields whose commutation relation reads as
\begin{equation}
\left[\frac{\partial_{r}\Theta(r_{1})}{\pi},\Phi\left(r_{2}\right)\right]=i\delta\left(r_{1}-r_{2}\right).
\end{equation}
The mapping between the spin operators of the microscopic model and the bosonic fields are as follows:
\begin{equation}
  \begin{split}
	&S^{x}(r)\sim\cos\left[\Phi(r)\right],\\
	&S^{y}(r)\sim-\sin\left[\Phi(r)\right],\\
    &S^{z}(r)\sim\partial_{r}\Theta(r)-C_{z}(-1)^{r}\sin\left[2\Theta(r)\right],\\
    &\vec{S}(r)\vec{S}(r+1)\sim(-1)^{r}\cos\left[2\Theta(r)\right].
  \end{split}
\end{equation}

Then for small $H_{\mathrm{int}}$, one can calculate the scaling dimension $\Delta$
of the two interaction terms~\cite{MotrunichTHE2019}:
\begin{equation}
  \begin{split}
	&\Delta[\cos(4\Theta)]=4g,\\
    &\Delta[\cos(2\Phi)]=1/g.
  \end{split}
\end{equation}
For $g>1/2$, the $\cos(4\Theta)$ term is relevent and $\cos(2\Phi)$
term is irrelevent. The ground state will be either a VBS (where $\Theta=0$ or
$\pi/2$) for $\lambda_{\Theta}<0$ or a Z-AFM (where $\Theta=\pi/4$ or
$3\pi/4$) for $\lambda_{\Theta}>0$. Likewise, for $g<1/2$, the
ground state will be either an $X$-FM for $\lambda_{\Phi}<0$ or a $Y$-FM
for $\lambda_{\Phi}>0$. Interestingly, at $g=1/2$, these two interaction terms are both
marginally irrelevant and the hamiltonian is symmetric under the $2\Theta\leftrightarrow\Phi$
interchange for $\lambda_{\Theta}=\lambda_{\Phi}<0$. Note that one may rewrite $\cos(4\Theta)$ to $\cos^{2}(2\Theta)-\sin^{2}(2\Theta)$,
and $\cos(2\Phi)$ to $\cos^{2}(\Phi)-\sin^{2}(\Phi)$. This means, at the $X$-FM
to VBS transition point (red line in Fig. \ref{fig:1}), there exists a continuously rotation between
the order parameters of the $X$-FM state, $\cos(\Phi)$, and the VBS state, $\cos(2\Theta)$.
Actually, there is an additional rotation between the order parameters of the $Y$-FM state ($\sin(\Phi)$) and
the $Z$-AFM state ($\sin(2\Theta)$). But a transition between the $Z$-AFM and VBS states would require $\lambda_{\Theta}=0$ and $g>1/2$, which cannot be satisfied in the original
microscopic model of Eq. \eqref{eq:spinmodel}.
Similar analysis can be applied
to the DQCP between the $Y$-FM and VBS (green line in Fig. \ref{fig:1})
for $\lambda_{\Theta}=-\lambda_{\Phi}<0$. Another transition takes place at $\lambda_{\Phi}=0$ and $g<1/2$, corresponding to the DQCP between the $X$-FM and $Y$-FM (blue line in Fig. \ref{fig:1}).

We investigate the enhanced symmetry at the tricritical point by approaching it along the isotropic line (fixing $J_{x}=J_{y}$ and increasing $K$ in Eq. \eqref{eq:spinmodel}, or equivalently, fixing $\lambda_{\Phi}=0$
and increasing $g$ in Eq. \eqref{eq:bosonization}). Along this line, the low-energy effective Hamiltonian is simplified to the standard sine-Gordon model. The scaling dimensions and critical behaviors of
these bosonic fields and corresponding spin operators can be calculated analytically.
It is well known that across the tricritical point from the Luttinger liquid phase (critical
line of the $X$-FM and $Y$-FM phases) to the VBS phase, the system will undergo a Kosterlitz-Thouless (KT) transition~\cite{Furusaki2012,fradkin2013BOOK,tsvelik2007BOOK,Affleck1998JPA,MotrunichTHE2019,SineGorScaling1975}.

In light of the scaling dimension analysis aforementioned, all the
interaction terms are irrelevant along the isotropic critical line. Therefore,
the Hamiltonian $H_{b}$ flows to $H_{0}$ under renormalization group transformations. It is noteworthy that
the $\cos(4\Theta)$ term at the tricritical point is marginally irrelevant,
which will cause a correction to scaling.

At the tricritical point, the effective sine-Gordon model can be rewritten in a fermionic representation defined on bond. The $\cos(4\Theta)$ term provides an irrelevant backscattering process corresponding to spin flip on bond\cite{fradkin2013BOOK,tsvelik2007BOOK,Affleck1998JPA,MotrunichTHE2019,SineGorScaling1975}. If we ignore the marginally irrelevant interaction, the model is mapped to
the $k=1$ $SU(2)_1$ Wess--Zumino--Novikov--Witten (WZNW) model\cite{tsvelik2007BOOK,Affleck1998JPA,fradkin2013BOOK,Sandvik2018Heisenberg}:
\begin{equation}
H_{\mathrm{WZNW}}=\int dx\left[\vec{J}_{L}\cdot\vec{J}_{L}+\vec{J}_{R}\cdot\vec{J}_{R}\right],
\end{equation}
where $\vec{J}_{L}$ and $\vec{J}_{R}$ are the left-move and right-move
current operators, respectively. They both satisfy the $\rm{SU}(2)_1$ Kac--Moody algebra.
Consequently, this WZNW model possessed two independent $\rm{SU}(2)$ symmetries.
However, the real situation is more complicated because the interaction term will mix the left-move
and right-move current. Note that in the infrared limit, the interaction
term is irrelevant. This means we should still obtain the WZNW model with two independent $SU(2)$ symmetries after a proper redefinition of the left- and right-moving modes. Therefore, we expect the symmetry at the tricritical point extends to $\rm{SU}(2)\times \rm{SU}(2)\sim {\rm O(4)}$, just like the case in the Heisenberg model\cite{tsvelik2007BOOK,Sandvik2018Heisenberg,Affleck1985PRL}. Instead of presenting an analytical proof, in the following, we justify this emergent $\rm O(4)$ symmetry by numerical calculations on the original microscopic model.

To examine the critical properties of the system, we calculate the correlation functions of the order parameters and the corresponding conserved currents. With the emergent symmetry, the order parameters of $X$-FM, $Y$-FM, $Z$-AFM,
and VBS form an ${\rm O(4)}$ pseudovector. In terms of microscopic spin operators, the components of the ${\rm O(4)}$ pseudovector read
\begin{equation}
  \begin{split}
	X\text{-FM} : &n_{1}(r_i)\sim S_{j}^{x}+S_{j+1}^{x}\\
	Y\text{-FM}: &n_{2}(r_i)\sim S_{j}^{y}+S_{j+1}^{y}\\
    Z\text{-AFM}: &n_{3}(r_i)\sim(-1)^{j}(S_{j}^{z}-S_{j+1}^{z})\\
    \text{VBS}: &n_{4}(r_i)\sim(-1)^{j}\vec{S}_{j}\vec{S}_{j+1}.
  \end{split}
\end{equation}
Moreover, the ${\rm O(4)}$ group has six Lie group generators which generates the
rotation between any two components of the ${\rm O(4)}$ pseudovector. As a consequence,
there must be 12 emergent conserved currents labeled as $\mathcal{J}_{ab}^{\mu}$,
corresponding to the continuous rotational symmetry of $n_{a}$ and
$n_{b}$ ($a,b=1,2,3,4$ and $a<b$) with space-time components labeled
by $\mu=\tau,r$. In terms of microscopic spin operators,
they can be written as
\begin{equation}
  \begin{split}
	&\mathcal{J}_{12}^{r}\sim\mathcal{J}_{34}^{\tau}\sim S_{j}^{z}+S_{j+1}^{z}\\
	&\mathcal{J}_{13}^{r}\sim\mathcal{J}_{24}^{\tau}\sim(-1)^{j}(S_{j}^{y}-S_{j+1}^{y})\\
    &\mathcal{J}_{23}^{r}\sim\mathcal{J}_{14}^{\tau}\sim(-1)^{j}(S_{j}^{x}-S_{j+1}^{x})\\
    &\mathcal{J}_{14}^{r}\sim\mathcal{J}_{23}^{\tau}\sim(-1)^{j}(S_{j}^{y}S_{j+1}^{z}+S_{j}^{z}S_{j+1}^{y})\\
    &\mathcal{J}_{24}^{r}\sim\mathcal{J}_{13}^{\tau}\sim(-1)^{j}(S_{j}^{x}S_{j+1}^{z}+S_{j}^{z}S_{j+1}^{x})\\
    &\mathcal{J}_{34}^{r}\sim\mathcal{J}_{12}^{\tau}\sim(S_{j}^{x}S_{j+1}^{y}-S_{j}^{y}S_{j+1}^{x}).
    \label{eq:currents}
  \end{split}
\end{equation}

The WZNW model predicts that the scaling dimensions of the conserved currents and
the vector components are $1$ and $1/2$ , respectively. The critical
exponent $\eta$ and the scaling dimension $\Delta$ are related by
$\eta=2\Delta$. However, due to the existence of marginally irrelevent
operators, the correlation functions of vector components acquire
a multiplicative logarithmic correction and behave as\cite{Affleck1998JPA,fradkin2013BOOK,tsvelik2007BOOK,Toshiya2017scaling}
\begin{equation}
  \begin{split}
	&\left\langle n_{a}(0)n_{a}(r)\right\rangle \sim\frac{\ln^{\frac{1}{2}}(r)}{r},a=1,2,4\\
	&\left\langle n_{3}(0)n_{3}(r)\right\rangle \sim\frac{f(r)}{r}.
    \label{eq:correlation}
  \end{split}
\end{equation}
Here, $f(r)$ is a weak correction term and would be unimportant for $r\gg1$\cite{Toshiya2017scaling}.
Note that the critical behaviour of the correlation functions of the vector components should be in accord with the prediction of the WZNW model only in the infrared limit. For the conserved currents, they should not acquire
such corrections to scaling. If the ${\rm O(4)}$ symmetry indeed emerges
at this point, the conservation law guarantees the scaling dimension
of the conserved currents to $\Delta=1$, \emph{i.e.} $\eta=2$ and the
correlation functions of conserved currents should behave as
\begin{equation}
    \left\langle \mathcal{J}_{ab}^{\mu}(0)\mathcal{J}_{ab}^{\mu}(r)\right\rangle \sim\frac{1}{r^2}.
\end{equation}
There are only 6 independent microscopic operators in Eq. \ref{eq:currents}
for the 12 conserved currents with $\mu=\tau,r$ and $a,b=1,2,3,4$. For
simplicity, we denote $\mathcal{J}_{ab}=\mathcal{J}_{ab}^{r}$ in the following discussion.
The corresponding components $\mathcal{J}_{ab}^{\tau}$ can be easily obtained
from Eq. \eqref{eq:currents}.

\section{Numerical Methods\label{sec:Numerical-Methods}}

In this work, we adopt a MPS based infinite time-evolving block decimation (iTEBD)\cite{Vidal,SCHOLLWOCK2011} method to study the ground-state
properties of the spin model defined in Eq. \eqref{eq:spinmodel}. Generally,
with sufficiently large Schmidt rank $D$, MPS can ensure accuracy
only for gapped systems. In the situation where the system is
gapless, with various well developed finite-$D$ or finite-entanglement
scaling techniques, MPS is capable of exploring the ground-state properties
trustingly\cite{scaling2008,scaling2009,scaling2012}. A finite-$D$
scaling technique will be detailed and applied in the analysis of
the KT transition.

A two-point correlator of operators $u$ and $v$ in the matrix product representation take the
generic form as\cite{SCHOLLWOCK2011}
\begin{equation}
\left\langle u(0)v(r)\right\rangle =\sum_{k=1}^{rank(D*D)}c_{k}(\Lambda_{k})^{r-1}.\label{eq:tensorco}
\end{equation}
Here, $\Lambda_{k}$s are the eigenvalues of the transfer operator
and $\Lambda_{1}$ is the largest eigenvalue. If we transform the
transfer operator into a normalized form with $\Lambda_{1}=1$, there
remains a constant term $c_{1}$ and an exponential decay term $\sum c_{k}\Lambda_{k}^{r-1}$
in Eq. \eqref{eq:tensorco}. The constant term guarantees the long range order.
The leading decay factor $\gamma=\Lambda_{2}$ (the second largest
eigenvalue) contributes an effective correlation length as
\begin{equation}
\xi=-1/\ln(\left|\gamma\right|).\label{eq:correlation_length}
\end{equation}
There are two parts in Eq. \eqref{eq:tensorco}. $(\Lambda_{k})^{r-1}$
terms reflect the intrinsic characteristic of the ground state, while
the $c_{k}$ terms dictate the specific correlation behaviors for
given operators of $u$ and $v$ in Eq. \eqref{eq:tensorco}. Also note that
the effective correlation length $\xi$ is obtained from the second largest eigenvalue, and hence is independent of the correlation function. Furthermore in the situation where the
system is gapless(actual correlation length $\tilde{\xi}\rightarrow\infty$),
the effective correlation length $\xi$ would conform to a power law
behavior as\cite{scaling2009}
\begin{equation}
\xi\sim D^{\kappa}.
\end{equation}
Here, $\kappa$ is the finite-entanglement scaling exponent related
to central charge $c$ as
\begin{equation}
\kappa=\frac{6}{c\left(\sqrt{\frac{12}{c}}+1\right)}.
\end{equation}
Instead, for the gapped system, correlation length is always a finite
value, even in the infinite-$D$ limit. Thus, the extrapolations of $\xi$ in the infinite-$D$ limit can make a distinction between gapped and gapless systems.

To determine a ground state in matrix product representation, one
usually starts from a random initialization. Deep inside an orderd
phase, the ground state is not sensitive to the way of initialization.
The local order parameters and ground-state energy are sufficiently
accurate with a finite $D$. However, when the system is in a quasi-long-range(QLR)
ordered phase, finite-$D$ effects may appear. The gap obtained in
a finite-$D$ MPS is always finite which leads to a finite value of
order parameters. This long range order is an artifact
and will disappear in the infinite-$D$ limit. For example of the ground-state
calculation on the isotropic line (blue line in Fig. \ref{fig:1})
in this work, the magnetic order parameters $\left\langle S^{x}\right\rangle $
and $\left\langle S^{y}\right\rangle $ are always finite values.
Moreover, the ${\rm U(1)}$ symmetry of Hamiltonian results in infinite
degenerate local minima of rotating between $\left\langle S^{x}\right\rangle $
and $\left\langle S^{y}\right\rangle $ with finite $D$. To eliminate
the effects of the finite values of order parameters as much as possible,
we fix the obtained magnetic order along one direction and calculate
the correlation function along vertical directions. For example, to
calculate the correlation function of $S^{x}$, we fix the magnetic
order along the $y$-direction on the isotropic line. It is noteworthy
that the effective correlation length $\xi$ is the intrinsic characteristic
of a wave function and will not be influenced by the direction of
the magnetic order.

\section{Numerical Analysis of the KT Transition\label{sec:Numerical-Analysis-of}}

\begin{figure}[h]
\includegraphics[width=1\textwidth]{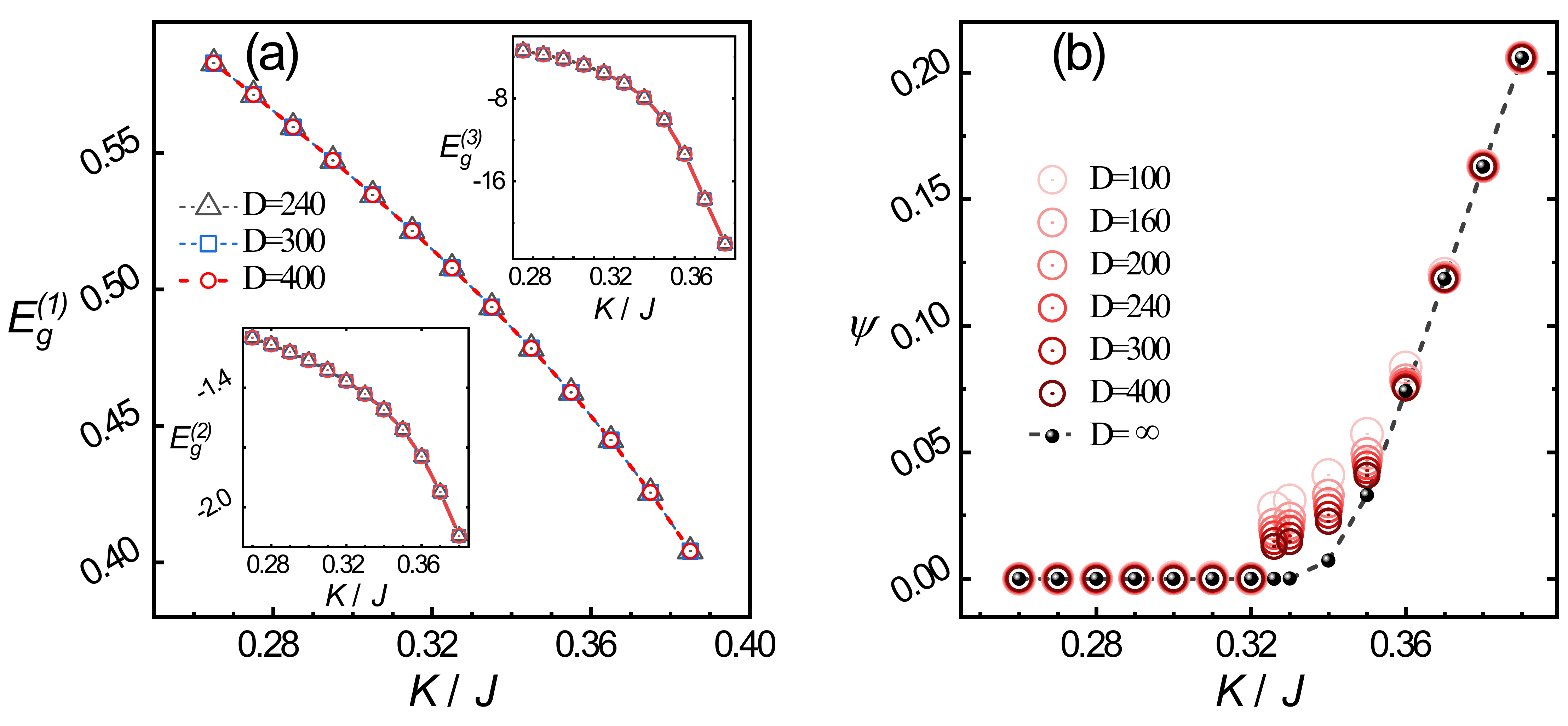}

\caption{(a) The first derivative $E_{g}^{(1)}$ of the ground-state energy
with $K/J$ along the blue line in Fig. \ref{fig:1}. The inserts
show the second derivative $E_{g}^{(2)}$ and third derivative $E_{g}^{(3)}$
of the ground-state energy, respectively. (b) The VBS order parameter
$\psi$ with $K/J$. \label{fig:2}}
\end{figure}

To accurately locate the location of the tricritical point, the most
convenient way is to approach it along the isotropic line($J_{x}=J_{y}$).
In this situation, the sine-Gordon model has predicted that the system
will undergo a KT transition from the Luttinger liquid phase to an
ordered phase. We first examine the behaviors of the Luttinger liquid
to VBS transition. For simplicity, we ignore the $x,y$ indices and
label $J_{x}=J_{y}$ as $J$ in the following sections.

The VBS order parameter $\psi=\sum_{j}(-1)^{j}\vec{S}_{j}\vec{S}_{j+1}$
has been calculated. As shown in Fig. \ref{fig:2}(b), $\psi$ exhibits
a near-continuous transition at the point about $K/J\thickapprox0.324$.
As the parameter $K/J$ approaches the transition point from the VBS
order phase, the split of $\psi$ for different $D$ becomes more obvious, which means a notable finite-$D$ effect. To eliminate
the finite-$D$ effect and obtain the behavior of order parameters
in the thermodynamic limit, we extrapolate their values in the large-$D$
limit, as also shown in Fig. \ref{fig:2}(b). The extrapolated results
of $\psi$ for $D=\infty$ exhibit a smooth and continuous transition.
We further calculate the ground-state energy $E_{g}$ , its first
derivative $E_{g}^{(1)}=\mathrm{\mathrm{d}}E_{g}/\mathrm{\mathrm{d}}K$,
the second derivative $E_{g}^{(2)}=\mathrm{\mathrm{d}}E_{g}^{(1)}/\mathrm{\mathrm{d}}K$
, and the third derivative $E_{g}^{(3)}=\mathrm{\mathrm{d}}E_{g}^{(2)}/\mathrm{\mathrm{d}}K$
to explore the transition. As demonstrated in Fig. \ref{fig:2}(a),
all these derivatives vary continuously across the transition, and
do not develop a singularity or discontinuity. This is consistent with a KT transition, which is an infinite-order transition.

\begin{figure}[h]
\includegraphics[width=1\textwidth]{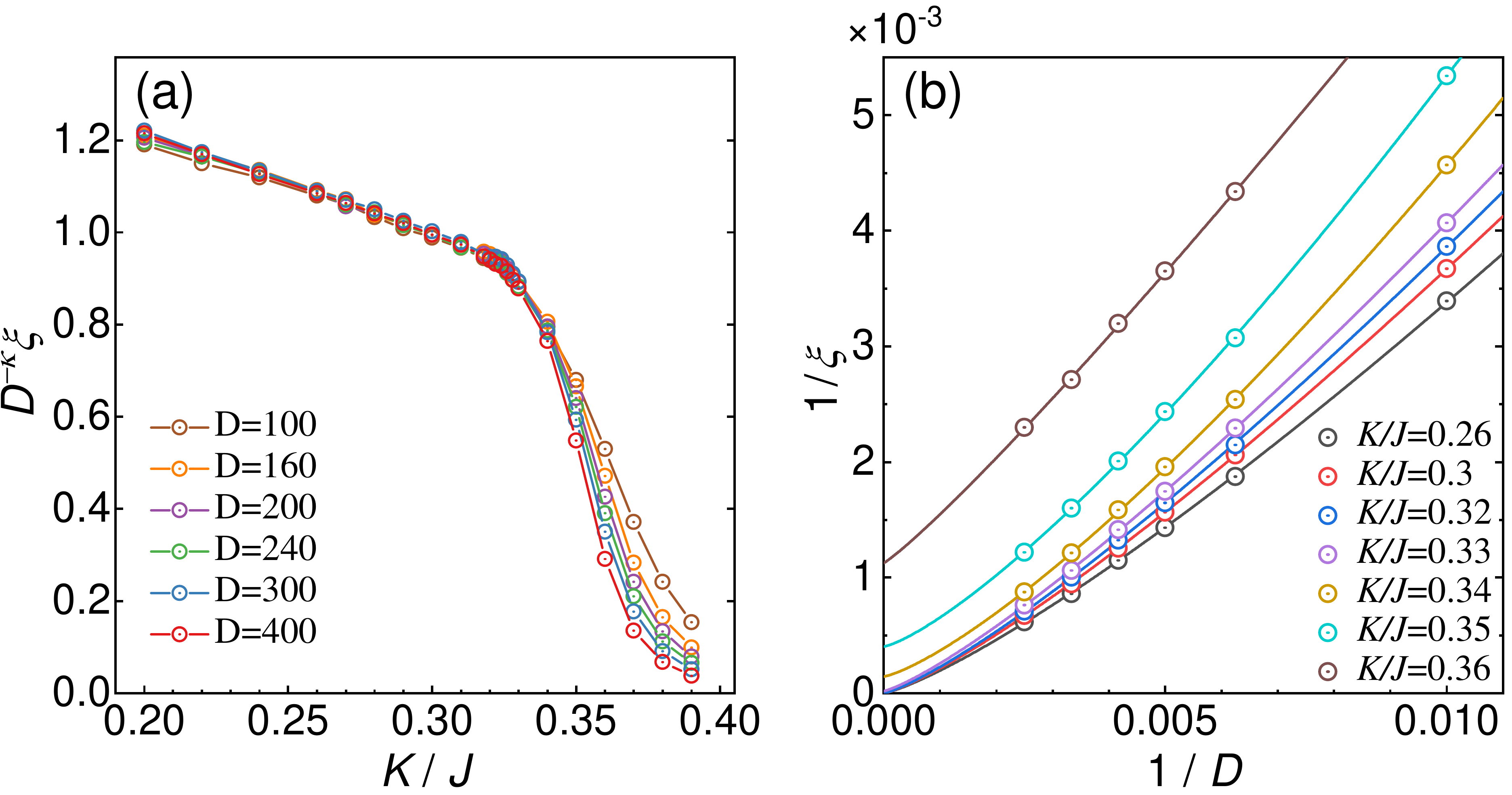}At the tricritical point, the
system is marginal to be a $n_{1}$ orderd, a $n_{2}$ orderd or a
$n_{3}$ phase.

\caption{(a) Scaling of the effective correlation length $\xi$ with $K/J$ across the
KT transition point along the blue line in Fig. \ref{fig:1} for different
$D$. (b) Scaling of $1/\xi$ as a function of $1/D$ for different
$K/J$. \label{fig:3}}
\end{figure}

To obtain more convincing evidence for the KT transition and
determine the exact location of the transition point, we calculate the
effective correlation length $\xi$ for different $D$. As demonstrated
in Fig. \ref{fig:3} (a), $\xi$ varies smoothly for small $K$, and
rapidly decay for $K/J\gtrsim0.33$, featuring a process from a gapless
system to a gapped system. The transition point can be determined by the kink to be $K_c/J\approx0.323\pm0.002$.
With properly choosing the exponent $\kappa\approx1.23$,
all $\xi$ data collapse when the system is gapless. The exponent $\kappa$ is a universal constant related to the central charge $c$ of the system. The $\kappa$ value we obtained is close to the predicted one, $\kappa\approx1.334$, for a system with central charge $c=1$.\cite{Huang2019,scaling2009}
Direct and clear results are also shown
in Fig. \ref{fig:3} (b). The results for $K/J\leq0.32$
can all be fitted by power law functions with $\kappa\approx1.23$
which is close to the predictive value of $1.344$. However, for the
gapped system, it is expected to observe a finite value of $\xi$
even in the $D=\infty$ limit. The results for $K/J\geq0.34$ in Fig.
\ref{fig:3} (b) can not be fitted by power functions without a constant
term and the extrapolations of $\xi$ for these results in the $D=\infty$
limit are all finite values. These results evidence a KT transition
for $0.32\lesssim K/J\lesssim0.33$.

\begin{figure}[h]
\includegraphics[width=1\textwidth]{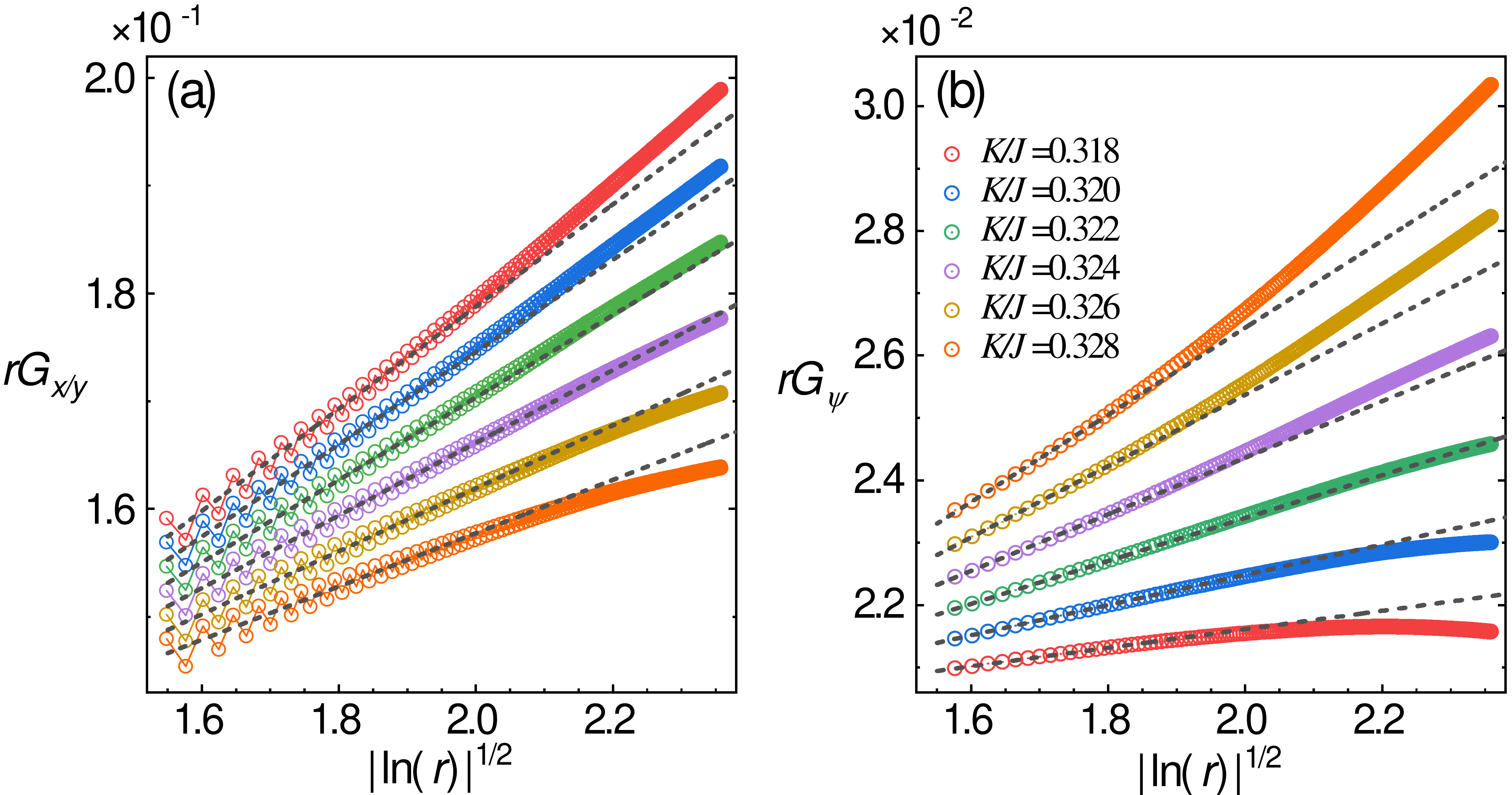}

\caption{(a)$rG_{x/y}$ and (b)$rG_{\psi}$ as a function of $\left|\ln(r)\right|^{1/2}$
for different $K/J$. $G_{x/y}$ and $G_{\psi}$ are the correlation
functions of $S_{x/y}$ and $\psi$ , respectively. The dashed linear
lines are guides to eyes. \label{fig:4}}
\end{figure}

The transition point can also be determined more accurately by another
method. At the tricritical point, the sine-Gordon model has predicted that
a multiplicative logarithmic correction term $\left|\ln(r)\right|^{1/2}$appears
in the correlation functions as mentioned in Eq. \ref{eq:correlation}.
As demonstrated in Fig. \ref{fig:4}, we plot $rG_{x/y}$ and $rG_{\psi}$
as a function of $\left|\ln(r)\right|^{1/2}$ for different $K/J$. Here $G_{u}=\langle u(0) u(r) \rangle -\langle u \rangle^2$.
Exactly at the tricritical point, both of these two plots will be linear.
The following work becomes to decide which one of the these curves
is closer to the perfect linear dependence. By combing the results
of Fig. \ref{fig:4} (a) and (b), we determine the transition
point is between $K/J=0.322$ to $K/J=0.324$. This result in consistent with above two methods, and it is also consistent with the estimates in previous works\cite{Furusaki2012,Nomura_1994}.

\section{Numerical Evidence of Emergent ${\rm O(4)}$ Symmetry\label{sec:Numerical-evidence-of}}

In this section we prove the existence of an enhanced
${\rm O(4)}$ symmetry by calculating the exponents $\eta$ of the $O(4)$ pseudovector components
related by the emergent symmetry and the associated emergent conserved
currents. There will be two sets of evidences for the existence of the enhanced
${\rm O(4)}$ symmetry. The first evidence is that the critical behaviors
and exponents $\eta$ of the vector components are all in accord with
the theoretical predictions of the WZNW model, which holds an $\rm SU(2)\times SU(2)\sim O(4)$ symmetry. The other robust evidence is that $\eta$ of all associated emergent conserved
currents are pinned to the integer value $2$.

\begin{figure}[h]
\includegraphics[width=1\textwidth]{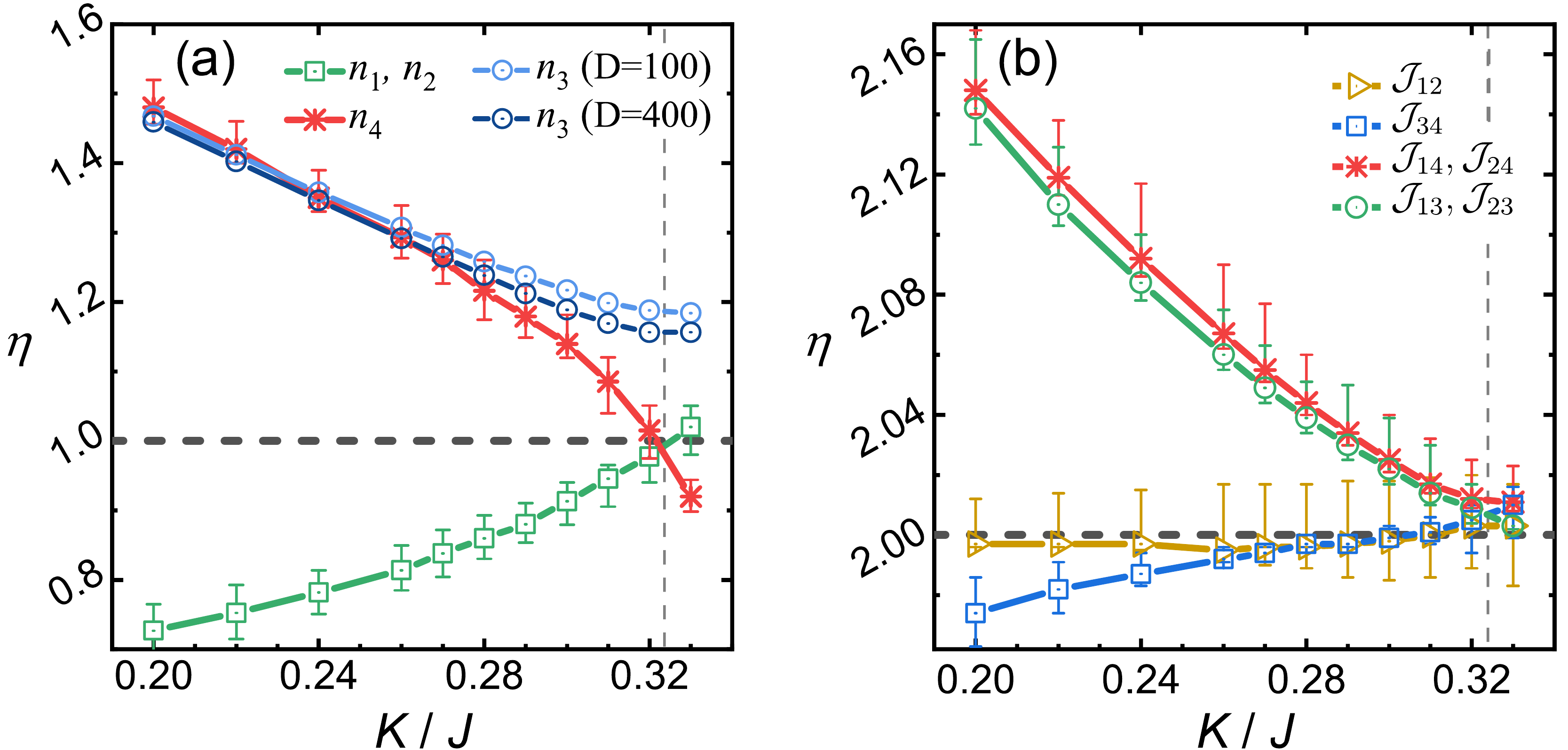}

\caption{The exponent $\eta$ of the $\rm O(4)$ pseudovector components (in (a) ) and associated conserved currents (in (b)) with $K/J$. Constrained by the LSM theorem, the ground state should
preserve the rotational symmetry between $S_{x}$($n_{1}$) and $S_{y}$($n_{2}$).
Hence, the results of interchanging the indices $1\leftrightarrow2$ are identical in the calculation. \label{fig:5}}
\end{figure}

The effective correlation length $\xi$ has been calculated in Sec
\ref{sec:Numerical-Analysis-of}. All $\xi$ values for the gapless system
are larger than $1400$ with $D=400$, and we fit the data up to about $r\lesssim1000$, which is reliable.

We first estimate $\eta$ of vector components by fitting correlation
functions for different $K/J$. As shown in Fig. \ref{fig:5}(a),
the $\eta$ of $n_{1}$, $n_{2}$, and $n_{4}$ at the tricritical point
are identical and close to the integer $1$. But the $\eta$ values of $n_{3}$ and $n_{4}$ differs when the system is close to the tricritical point. By comparing the $D=400$ data with those of $D=100$, we clearly show that this is
a finite-$D$ effect, which lies in the microscopic model because the marginal irrelevance only strictly holds in the thermodynamic limit (infrared limit). For finite-$D$ systems, corrections that are relevant always exist. At the tricritical point,
the system is marginal to be in either a $n_{1}$ ordered, or a $n_{2}$ ordered,
or a $n_{3}$ ordered phase. Therefore, $n_{4}$ is more \emph{relevant}
than $n_{3}$ in short range. Nevertheless, in the infrared limit,
this two $\eta$ will both flow to $1$, as suggested by the finite-$D$ study in Fig.~\ref{fig:5} (a).

The correlation-function exponents of the emergent
conserved currents are also estimated. As demonstrated in Fig. \ref{fig:5}(b),
the evidences from $\eta$ of conserved currents are more robust.
The $\eta$ of all conserved currents at the tricritical points are identical
to be $2$. Another interesting result shown in Fig. \ref{fig:5}(b)
is that the $\eta$ approximately keeps symmetric for $\mathcal{J}_{12}\leftrightarrow\mathcal{J}_{34}$,
$\mathcal{J}_{13}\leftrightarrow\mathcal{J}_{24}$ , and $\mathcal{J}_{14}\leftrightarrow\mathcal{J}_{23}$
interchange. It reflects a space-time duality of the conserved currents.

\begin{figure}[h]
\includegraphics[width=1\textwidth]{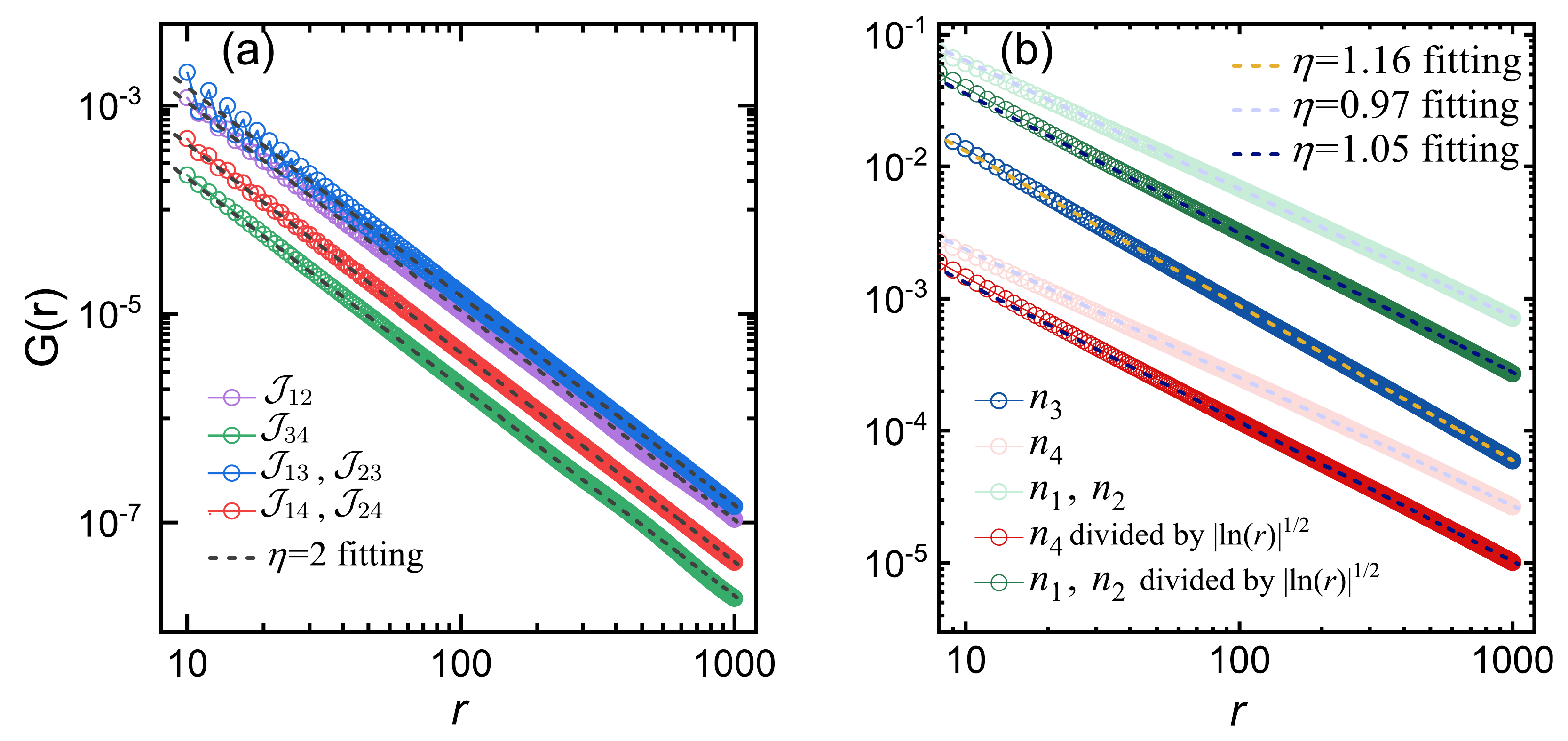}

\caption{Correlation functions of (a) emergent conserved currents and (b) vector
components at the tricritical point.\label{fig:6}}
\end{figure}

To further show the evidence of the emergent ${\rm O(4)}$ symmetry clearly,
we calculate the correlation functions of vector components and associated
emergent conserved currents just at the tricritical point($J_{x}=J_{y}$,
$K/J=0.322$) as shown in Fig. \ref{fig:6}. The dashed lines in Fig.
\ref{fig:6}(a) are guide lines with $\eta=2$. All these emergent
conserved currents are in accord with the expected behaviors of $\eta=2$
very well. The exponents of these 4 vector components in Fig.
\ref{fig:6}(b) are approximately equal, signaling the $\rm O(4)$ symmetry.

\section{Discussions and Conclusions\label{sec:Discussions and Conclusions}}

As mentioned in the introduction, the origin of emergent enhanced
continuous symmetry at the transition point is still challenging. In this work, we systematically investigate the enhanced ${\rm O(4)}$
symmetry on a 1D tricritical point. We justify the emergence of this symmetry by combining analytical analysis with numerical calculations. Note that the exact symmetry only exist in the infrared limit, which is usually difficult for numerics. To address this issue numerically, employ the iTEBD method to study the ground-state
properties of the system.
We verify the emergent symmetry
by two sets of evidences: checking the critical exponents with the
symmetry-enhanced low-energy effective theory, and verifying if the
scaling dimensions of conserved currents are pinned to an integer. The agreement between numerical and analytical results indicates the numerical method is capable of handing this issue. Application of our method to other problems would be interesting. Note that our method also provides a new approach to study the
KT transition numerically.

For this specific model, the marginally irrelevant terms result in
that the system only keep the enhanced ${\rm O(4)}$ symmetry in the infrared
limit. But an extended $\rm{SU}(2)\times {\rm U(1)}$ symmetry still preserves
in the short length scale. As an extension of this work, if we consider
a global phase diagram with an extra tuning parameter, such as an
AFM exchange coupling along $S^{z}$, enhanced ${\rm O(4)}$ symmetry at
some transition points may be more general.

In higher dimensions, verifying the emergent symmetry by conserved
currents at DQCPs has been applied to some systems\cite{Meng2018,Meng2019}.
Besides a deconfined tansition, emergent symmetry in higher dimensions may be even of first order, where the emergent Goldstone mode associated with the enhanced symmetry appears at the transition. In this case, the system keeps gapless at the first-order transition, and the analysis
based on exponents of correlation functions in our method will still be applicable.

\section*{References}{}

\end{document}